\documentclass[]{raa}            
\usepackage{graphicx,times}
\usepackage{natbib}

\providecommand{\bm}[1]{\mbox{\boldmath$#1$\unboldmath}}

\begin{document}
\title{A revised graduated cylindrical shell model and its application to a prominence eruption}

\volnopage{ {\bf --} Vol.\ {\bf --} No. {\bf XX}, 000--000}
   \setcounter{page}{1}

\author{Qing-Min~Zhang\inst{1,2}, Zhen-Yong~Hou\inst{3}, Xian-Yong~Bai\inst{4,5,6}}

\institute{$^1$Key Laboratory of Dark Matter and Space Astronomy, Purple Mountain Observatory, Nanjing 210023, China; {\it zhangqm@pmo.ac.cn} \\
           $^2$Yunnan Key Laboratory of Solar physics and Space Science, Kunming 650216, China \\
           $^3$School of Earth and Space Sciences, Peking University, Beijing 100871, China \\
           $^4$National Astronomical Observatories, Chinese Academy of Sciences, Beijing 100101, China \\
           $^5$University of Chinese Academy of Sciences, Beijing 100049, China \\
           $^6$Key Laboratory of Solar Activity and Space Weather, National Space Science Center, Chinese Academy of Sciences, Beijing 100190, China \\
 }

\abstract{In this paper, the well-known graduated cylindrical shell (GCS) model is slightly revised by introducing longitudinal and latitudinal deflections of prominences originating from active regions (ARs).
Subsequently, it is applied to the three-dimensional (3D) reconstruction of an eruptive prominence in AR 13110, 
which produced an M1.7 class flare and a fast coronal mass ejection (CME) on 2022 September 23.
It is revealed that the prominence undergoes acceleration from $\sim$246 to $\sim$708 km s$^{-1}$. 
Meanwhile, the prominence experiences southward deflection by 15$\degr$$\pm$1$\degr$ without longitudinal deflection, suggesting that the prominence erupts non-radially.
Southward deflections of the prominence and associated CME are consistent, validating the results of fitting using the revised GCS model.
Besides, the true speed of the CME is calculated to be 1637$\pm$15 km s$^{-1}$, which is $\sim$2.3 times higher than that of prominence.
This is indicative of continuing acceleration of the prominence during which flare magnetic reconnection reaches maximum beneath the erupting prominence.
Hence, the reconstruction using the revised GCS model could successfully track a prominence in its early phase of evolution, including acceleration and deflection.
}

\authorrunning{Zhang et al.}
\titlerunning{A revised GCS model}
\maketitle

\section{Introduction} \label{sect:intro}
Solar flares and coronal mass ejections (CMEs) are the most powerful activities in the solar atmosphere, which have drastic and profound influences on the heliosphere (\citealt{chen11, sm11, re13}).
The primary origins of flares and CMEs are believed to be impulsive eruptions of solar prominences or filaments (\citealt{jan15}).
Prominences observed in H$\alpha$ or extreme-ultraviolet (EUV) wavelengths usually show helical structures (\citealt{kum12}),
and fast rotations or untwisting motions are frequently detected during eruptions \citealt{gre07, yan14, shen19, zhou23}.
Before loss of equilibrium, the gravity of a prominence is balanced by the upward tension force of magnetic dips within a sheared arcade or a flux rope (\citealt{liu12, chen18, zhou18, luna21, guo22}).
A magnetic flux rope comprises a bundle of twisted field lines, which are wrapping around a common axis (\citealt{td99, qiu04, wang15, gou23}).
Flux ropes play a central role in driving flares and CMEs (\citealt{am03, rou03, au10, cx13, in18, mei20, jia21}).
Sometimes, they could be heated up to $\sim$10 MK before or during eruptions and are termed as hot channels (\citealt{zj12, cx13, liu22, zqm22b}), 
which are merely observed in 94 {\AA} and 131 {\AA} of the Atmospheric Imaging Assembly (AIA; \citealt{lem12}) on board the Solar Dynamics Observatory (SDO) spacecraft.
Flux ropes propagate radially in most cases. However, a fraction of them undergo deflections and propagate non-radially (\citealt{guo19, mj19, hess20, zqm22a}).
The inclination angle with the normal direction lies in the range of 15$\degr$-70$\degr$.
In the typical three-part structure of CMEs, the dark cavity and bright core are considered to be a flux rope and the embedded prominence (\citealt{ih85, song23}).

The three-dimensional (3D) shape and direction of a CME are essential in estimating the arrival time and geo-effectiveness of a CME.
The well-known cone model, resembling an ice cream, was proposed and applied to investigate the evolutions of morphology and kinematics of halo CMEs (\citealt{mich03, xie04}).
This model assumes a constant angular width and a constant linear speed during propagation in radial direction \citep{zqm10}.
Considering that a part of prominences and the driven CMEs propagate non-radially, \citet{zqm2021} put forward a revised cone model and applied it to two prominence eruptions.
The tip of the cone is located at the source region of CME. The model is characterized by four parameters: the length ($r$) and angular width ($\omega$) of the cone,
and two angles ($\phi_1$ and $\theta_1$) denoting the deflections in longitudinal and latitudinal directions. 
Using this model, \citet{zqm2022} satisfactorily tracked the 3D evolution of a halo CME as far as $\sim$12 $R_\odot$ on 2011 June 21.

\citet{the06} proposed the graduated cylindrical shell (GCS) model to perform 3D reconstructions of flux rope-like CMEs (\citealt{vour13}). 
The flux rope in their model looks like a croissant, which has two identical legs with a length of $h$ and angular separation of 2$\alpha$ (\citealt{the09, the11}).
The legs are connected by a circulus with varying cross sections so that the aspect ratio $\kappa$ keeps constant.
Another angle $\gamma$ represents the tilt angle of the polarity inversion line (PIL) of the source region with a longitude $\phi$ and a latitude $\theta$, respectively.
Besides, electron number density ($N_e$) is considered to  synthesize white-light (WL) images observed by coronagraphs.
Thanks to multiperspective observations from the Large Angle and Spectrometric Coronagraph (LASCO; \citealt{bru95}) on board the SOHO spacecraft
and WL coronagraphs (COR1, COR2) on board the twin Solar TErrestrial RElations Observatory (STEREO; \citealt{kai08}) spacecraft,
the GCS model has been widely used to perform 3D reconstructions of CMEs (\citealt{mi09, cx14, mo14, lie15, lu17, sa23, zhou23}).
\citet{is16} developed an analytic 3D model for flux rope-like CMEs that incorporate all major deformations during their propagations, 
such as deflection, rotation, ``pancaking", front flattening, and skewing.

The 3D morphologies of eruptive prominences could be obtained using the triangulation technique 
when simultaneous observations from two or three perspectives are available (\citealt{tho09, li11, bi13, guo19}).
Deflection, kinking, and rotation of the prominences are found based on the 3D reconstruction. 
Until now, the GCS model has rarely been applied to the reconstruction of eruptive prominences, especially those propagating non-radially.
In this paper, the GCS model is slightly modified and applied to reconstruct the shapes of an eruptive prominence in NOAA active region (AR) 13110 (N16E84), 
which produced a GOES M1.7 class flare and a fast CME on 2022 September 23.
The model is described in Section \ref{sect:model}. The results of 3D reconstruction are presented in Section \ref{sect:data}.
A brief summary and discussions are given in Section \ref{sect:sum}.

\section{Revised GCS model}
\label{sect:model}

Similar to the revised cone model, the GCS model is also modified in two aspects: 
Firstly, the tip of the two legs is located at the source region of the eruptive prominence rather than the solar center.
This applies to flux ropes originating from active regions, instead of quiescent prominences with much longer extensions (\citealt{li11, dai21, zhou23}).
It should be emphasized that the footpoints of a flux rope have separation and are not strictly close to each other \citep{wang15}.
Moreover, the footpoints may experience long-distance migration during eruption \citep{gou23}. In this respect, the assumption that the footpoints of a flux rope are cospatial is relatively strong.
Secondly, the GCS symmetry axis passing through the circulus has inclination angles of $\phi_1$ and $\theta_1$ with respect to the local longitude and latitude, respectively.
The parameters $h$, $\alpha$, $\kappa$, $\gamma$, $\phi$, and $\theta$ have the same meanings (\citealt{the06}).
$\gamma=0\degr$ and $\gamma=90\degr$ indicate that the PIL is parallel and perpendicular to the longitude, respectively.
Since the traditional GCS model reduces to the ice-cream cone model when $\alpha=0$ (\citealt{the09}), the revised GCS model also reduces to the revised cone model when $\alpha=0$ \citep{zqm2021}.

The transform between the heliocentric coordinate system (HCS; $X_h$, $Y_h$, $Z_h$) and local coordinate system (LCS; $X_l$, $Y_l$, $Z_l$) is (\citealt{zqm2022}):
\begin{equation} \label{eqn-1}
\left(
\begin{array}{c}
x_h  \\
y_h  \\
z_h \\
\end{array}
\right)
=M_2
\left(
\begin{array}{c}
x_l  \\
y_l  \\
z_l \\
\end{array}
\right)
+
\left(
\begin{array}{c}
R_{\odot}\cos{\theta}\cos{\phi} \\
R_{\odot}\cos{\theta}\sin{\phi} \\
R_{\odot}\sin{\theta} \\
\end{array}
\right),
\end{equation}
where
\begin{equation} \label{eqn-2}
M_2=
\left(
\begin{array}{ccc}
\sin{\theta}\cos{\phi} & -\sin{\phi} & \cos{\phi}\cos{\theta}  \\
\sin{\theta}\sin{\phi}  &  \cos{\phi} & \sin{\phi}\cos{\theta} \\
-\cos{\theta}                    &         0          &  \sin{\theta} \\
\end{array}
\right).
\end{equation}

The transform between LCS and GCS flux-rope coordinate system (FCS; $X_f$, $Y_f$, $Z_f$) is:

\begin{equation} \label{eqn-3}
\left(
\begin{array}{c}
x_l  \\
y_l  \\
z_l \\
\end{array}
\right)
=M_1
\left(
\begin{array}{c}
x_f  \\
y_f  \\
z_f \\
\end{array}
\right),
\end{equation}
where
\begin{equation} \label{eqn-4}
M_1=
\left(
\begin{array}{ccc}
\cos{\theta_1}\cos{\phi_1} & -\sin{\phi_1} & \cos{\phi_1}\sin{\theta_1} \\
\cos{\theta_1}\sin{\phi_1}  &  \cos{\phi_1} & \sin{\phi_1}\sin{\theta_1} \\
-\sin{\theta_1}                    &         0          &  \cos{\theta_1} \\
\end{array}
\right).
\end{equation}

   \begin{figure}
   \centering
   \includegraphics[width=0.8\textwidth, angle=0]{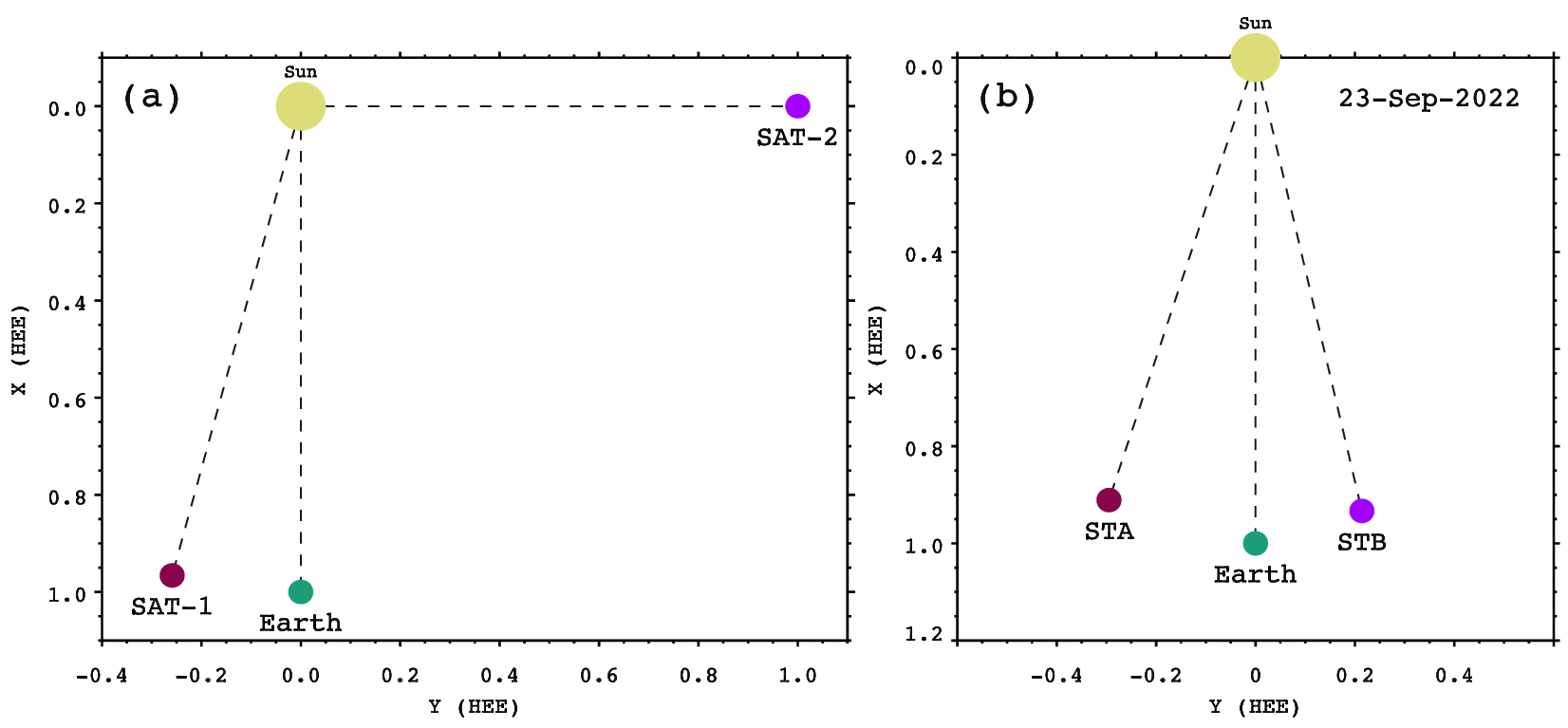}
   \caption{(a) Positions of Earth (green circle) and two artificial satellites. 
   SAT-1 (maroon circle) and SAT-2 (purple circle) have separation angles of -15$\degr$ and 90$\degr$ with the Sun-Earth connection, respectively.
   (b) Positions of Earth (green circle), ahead STEREO (STA, maroon circle), and behind STEREO (STB, purple circle) on 2022 September 23.}
   \label{fig1}
   \end{figure}

To reconstruct the shape of a flux rope in the revised model, observations from multiple viewpoints are needed as far as possible. 
In Figure~\ref{fig1}(a), the relative positions of Earth and two artificial satellites (SAT-1 and SAT-2) are denoted with green, maroon, and purple circles, respectively.
The separation angles between the artificial satellites with the Sun-Earth connection are denoted by $\xi_1$ and $\xi_2$, respectively. 
Note that SAT-1 and SAT-2 could be the ahead STEREO (hereafter STA) and behind STEREO (hereafter STB), 
or Extreme Ultraviolet Imager (EUI; \citealt{roch20}) on board Solar Orbiter (SolO; \citealt{mu20}), 
or Wide-Field Imager for Solar Probe Plus (WISPR; \citealt{vour16}) on board Parker Solar Probe (PSP; \citealt{fox16}).
Note that both SolO and PSP are much closer to the Sun than STEREO.
Consequently, the transform between the SAT-1 coordinate system ($X_{s1}$, $Y_{s1}$, $Z_{s1}$) and HCS is:

\begin{equation} \label{eqn-5}
\left(
\begin{array}{c}
x_{s1}  \\
y_{s1}  \\
z_{s1} \\
\end{array}
\right)
=M_{s1}
\left(
\begin{array}{c}
x_h  \\
y_h  \\
z_h \\
\end{array}
\right),
\end{equation}
where
\begin{equation} \label{eqn-6}
M_{s1}=
\left(
\begin{array}{ccc}
\cos{\xi_{1}} & \sin{\xi_{1}} & 0 \\
-\sin{\xi_{1}} & \cos{\xi_{1}} & 0 \\
0                  &     0              & 1 \\
\end{array}
\right).
\end{equation}

Similarly, the transform between the SAT-2 coordinate system ($X_{s2}$, $Y_{s2}$, $Z_{s2}$) and HCS is:

\begin{equation} \label{eqn-7}
\left(
\begin{array}{c}
x_{s2}  \\
y_{s2}  \\
z_{s2} \\
\end{array}
\right)
=M_{s2}
\left(
\begin{array}{c}
x_h  \\
y_h  \\
z_h \\
\end{array}
\right),
\end{equation}
where
\begin{equation} \label{eqn-8}
M_{s2}=
\left(
\begin{array}{ccc}
\cos{\xi_{2}} & \sin{\xi_{2}} & 0 \\
-\sin{\xi_{2}} & \cos{\xi_{2}} & 0 \\
0                  &     0              & 1 \\
\end{array}
\right).
\end{equation}

   \begin{figure}
   \centering
   \includegraphics[width=0.8\textwidth, angle=0]{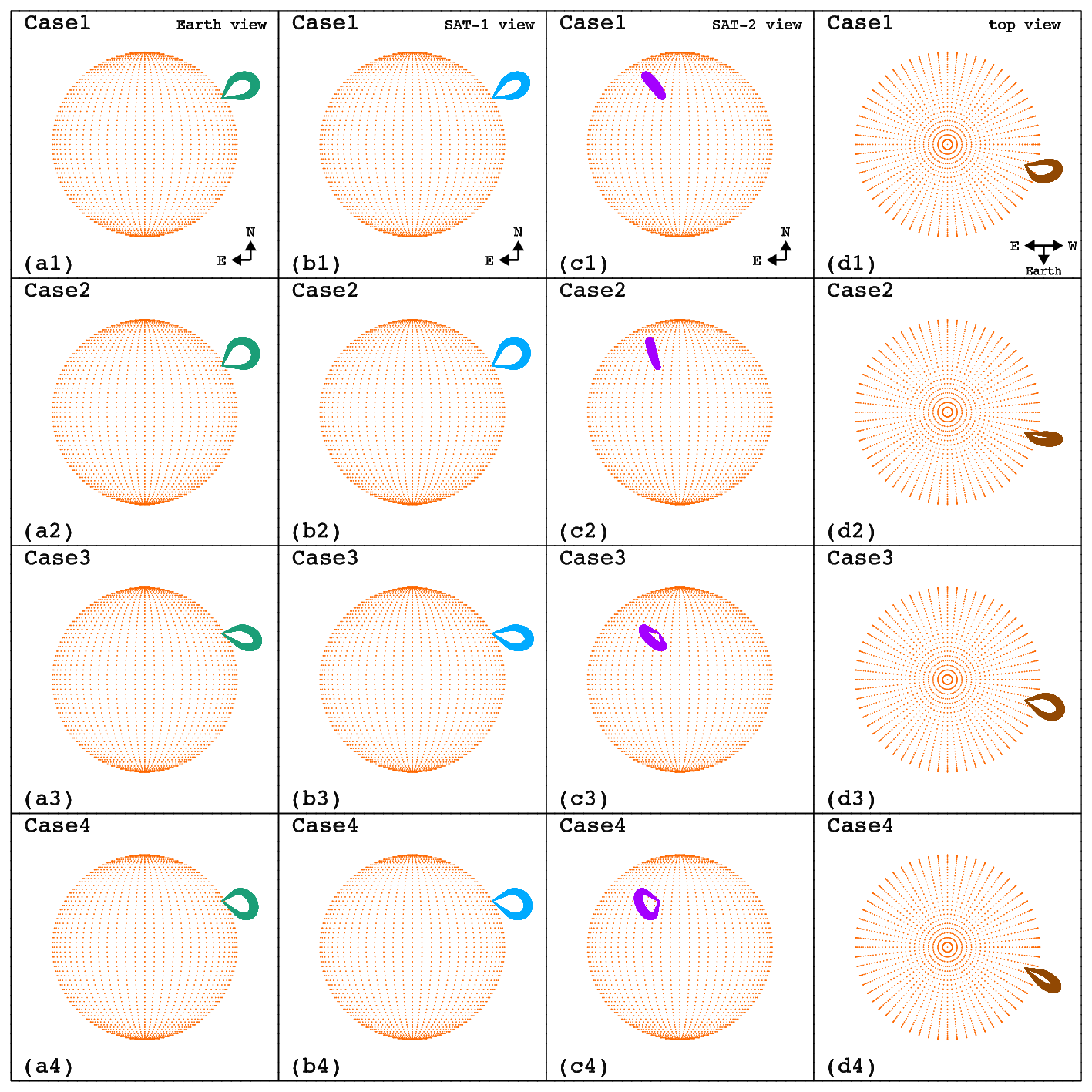}
   \caption{Different views of four artificial flux ropes (Case1$-$Case4) in the revised GCS model, see text for details.}
   \label{fig2}
   \end{figure}

To show the revised GCS model more clearly, four artificial flux ropes are created, in which $h=200\arcsec$, $\alpha=35\degr$, $\kappa=0.1045$, $\delta=\arcsin(\kappa)=6\degr$. 
The source region is characterized by $\phi=75\degr$, $\theta=30\degr$, and $\gamma=45\degr$.
The differences between the four flux ropes lie in the inclination angles ($\phi_1$ and $\theta_1$) of the symmetry axis, which are listed in Table~\ref{tab-1}.
In Case1, the direction of flux rope axis is exactly radial and there is no deflection. In Case2 (Case3), the flux rope experiences longitudinal (latitudinal) deflection, respectively. 
In Case4, there are deflections in both directions. Take $\xi_1=-15\degr$ and $\xi_2=90\degr$ (see Figure~\ref{fig1}(a)), 
Figure~\ref{fig2} shows different views of four flux ropes from Earth (first column), SAT-1 (second column), SAT-2 (third column), and solar North Pole (last column).
SAT-1 has a smaller separation angle with Earth, so that the morphologies of flux ropes from these two perspectives have slight differences.
Since SAT-2 is orthogonal to Earth, the morphologies of flux ropes from these two perspectives represent face-on and edge-on views, respectively.
In next Section, the revised GCS model will be applied to an eruptive prominence on 2022 September 23 without considering the electron number density.

\begin{table}
\begin{center}
\caption[]{Parameters of $\phi_1$ and $\theta_1$ in four Cases.}\label{tab-1}
 \begin{tabular}{c|c|c|c}
  \hline\noalign{\smallskip}
Case &   $\phi_1$     & $\theta_1$ & Deflection   \\
         &   (degree)    &  (degree)    &                    \\
  \hline\noalign{\smallskip}
Case1 & 0 & 0 & no \\
Case2 & -25 & 0 & longitudinal \\
Case3 & 0 & 40 & latitudinal \\
Case4 & -25 & 40 & both \\
  \noalign{\smallskip}\hline
\end{tabular}
\end{center}
\end{table}

\section{Application to a Prominence Eruption}
\label{sect:data}

\subsection{Flare and CME}

\begin{figure}
   \centering
   \includegraphics[width=0.7\textwidth, angle=0]{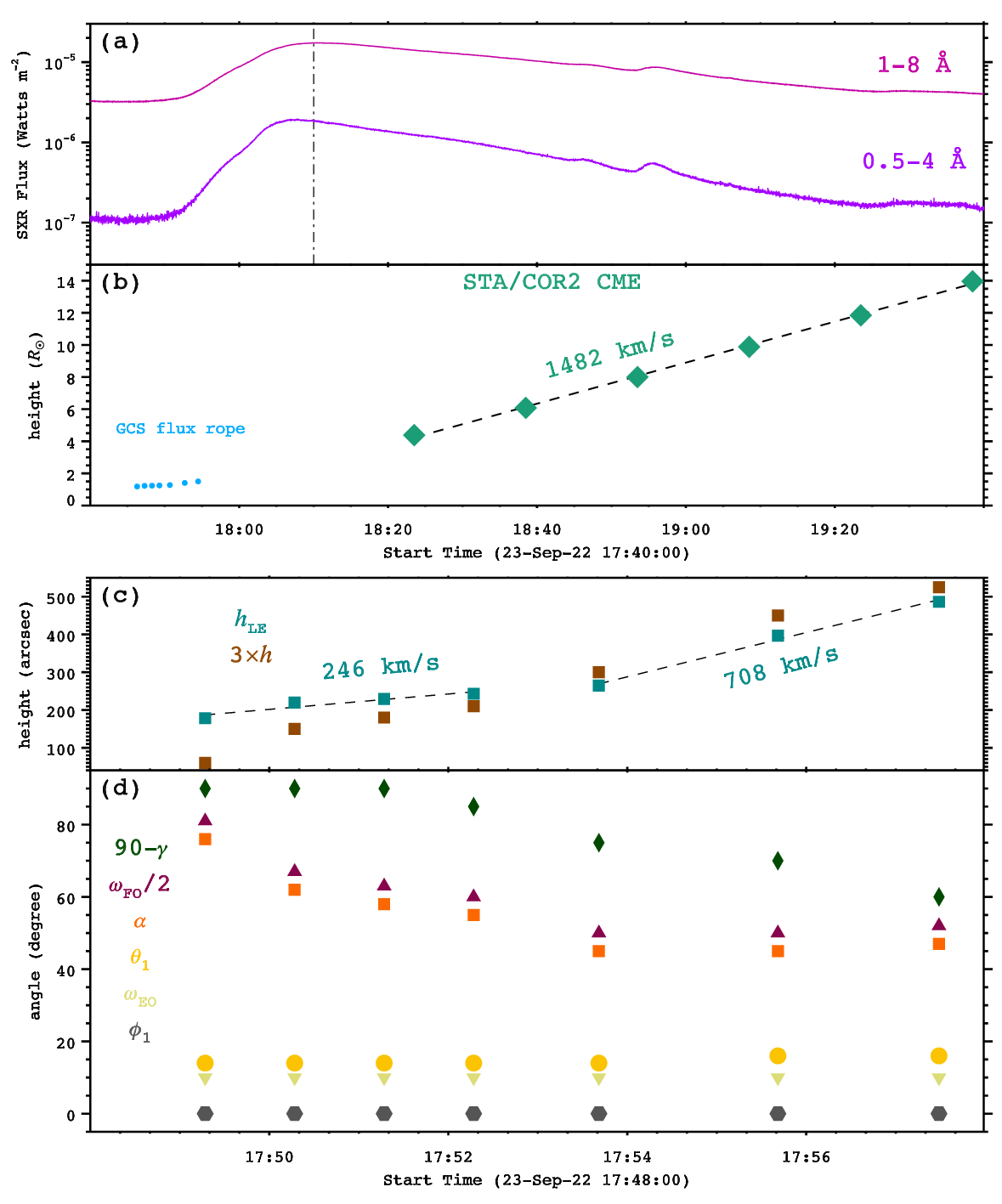}
   \caption{(a) GOES SXR light curves of the M1.7 flare in 1$-$8 {\AA} (red line) and 0.5$-$4 {\AA} (purple line). The dash-dotted line marks the peak time (18:10:00 UT).
   (b) Height-time plots of the leading edges of the reconstructed flux rope (blue circles) and CME observed by STA/COR2 (green diamonds).
   (c) Height-time plots of 3$h$ (brown squares) and $h_{\mathrm{LE}}$ (dark cyan squares). 
   Linear fittings of $h_{\mathrm{LE}}$ are performed before and after 17:53:00 UT, with the speeds being labeled.
   (d) Time variations of the fitted parameters, including $90-\gamma$ (green rhombuses), $\omega_{\mathrm{FO}}/2$ (purple triangles),
   $\alpha$ (orange squares), $\theta_1$ (yellow circles), $\omega_{\mathrm{EO}}$ (yellow triangles), and $\phi_1$ (gray hexagons), respectively.}
   \label{fig3}
   \end{figure}

\begin{figure}
   \centering
   \includegraphics[width=0.9\textwidth, angle=0]{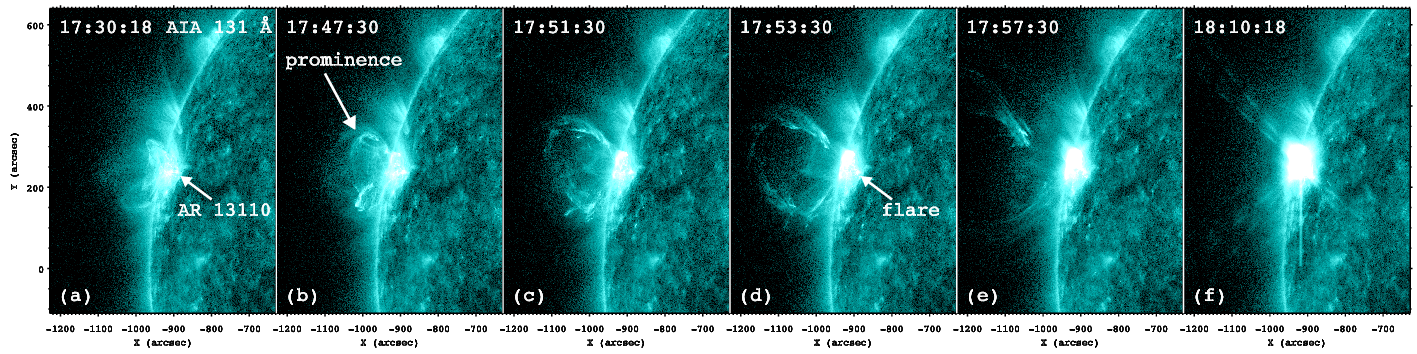}
   \caption{AIA 131 {\AA} images to illustrate the evolutions of the prominence and flare. The white arrows point to AR 13110, eruptive prominence, and hot flare loops.
   An animation showing the flare and prominence eruption in AIA 131 {\AA} is available.
   It covers a duration of 50 minutes from 17:30 UT to 18:20 UT on 2022 September 23. The entire movie runs for 6 s.
   (An animation of this figure is available.)}
   \label{fig4}
   \end{figure}

The event occurred in AR 13110, accompanied by an M1.7 flare and a fast CME. Figure~\ref{fig3}(a) shows SXR light curves of the flare in 1$-$8 {\AA} (red line) and 0.5$-$4 {\AA} (purple line).
The SXR emissions increase from 17:48:00 UT, peak at 18:10:00 UT, and decrease slowly until $\sim$18:50:00 UT.
Time evolutions of the prominence eruption and flare are illustrated by six 131 {\AA} images observed by SDO/AIA in Figure~\ref{fig4} and the associated online movie (\textit{anim131.mp4}).
Panel (a) shows AR 13110 with weak brightening before eruption. The prominence shows up and stands out after $\sim$17:46:00 UT (panel (b)).
It continues to rise and expands in height, during which the flare loops brighten significantly (panels (c-d)). 
The prominence accelerates and the apex escapes the field of view (FOV) of AIA, leaving behind the hot post-flare loops that cool down gradually (panels (e-f)).
It is noticed that the footpoints of the prominence remain in the AR without considerable separation.
The morphological evolution of the prominence is similar in other EUV and 1600 {\AA} wavelengths of AIA, indicating its multithermal nature (\citealt{li22a, zqm22a}).

\begin{figure}
   \centering
   \includegraphics[width=0.7\textwidth, angle=0]{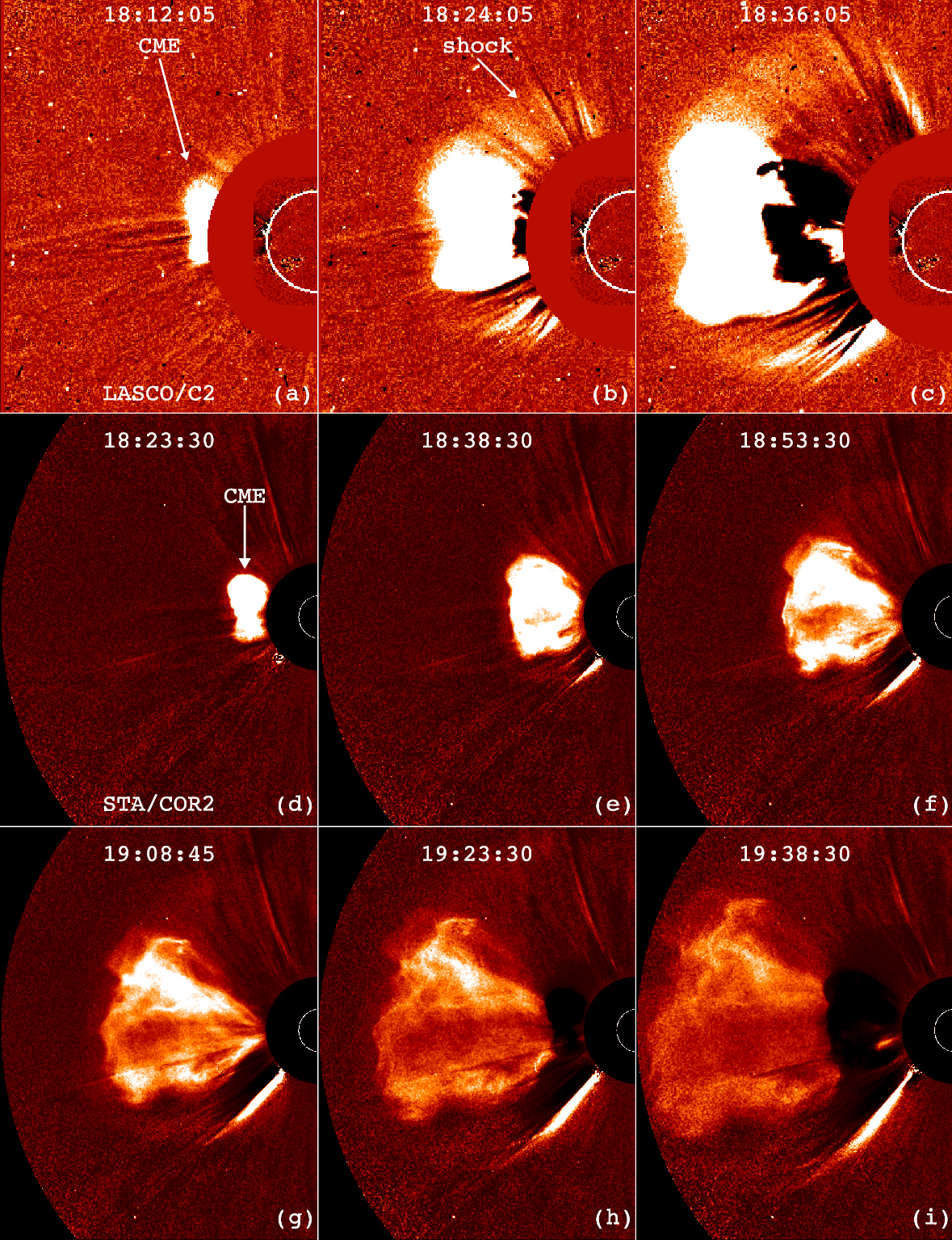}
   \caption{(a-c) Running-difference images of the related CME observed by LASCO/C2 during 18:12$-$18:36 UT.
   (d-i) Running-difference images of the CME observed by STA/COR2.
   The arrows point to the CME that first appears in the coronagraphs.}
   \label{fig5}
   \end{figure}

In Figure~\ref{fig5}, the top panels show running-difference WL images of the related CME observed by LASCO/C2. 
The CME\footnote{www.sidc.be/cactus/} first appears at 18:12:00 UT and propagates eastward with an angular width of $\sim$50$\degr$ 
and at a speed of $\sim$1644 km s$^{-1}$ (see Table~\ref{tab-2}). It is worth mentioning that the angular width is measured for the CME itself.
Since an interplanetary shock wave was driven by the CME (Figure ~\ref{fig5}(b-c)), the recorded angular width of the CME reaches 189$\degr$, 
which is much wider than the CME itself \footnote{cdaw.gsfc.nasa.gov/CME\_list/UNIVERSAL\_ver1/2022\_09/univ2022\_09.html}.
In Figure~\ref{fig1}(b), the green, maroon, and purple circles represent the positions of Earth, STA, and STB on 2022 September 23.
The twin satellites had separation angles of -17.9$\degr$ and 12.9$\degr$ with the Sun-Earth connection, although STB stopped working after 2016.
The middle and bottom panels of Figure~\ref{fig5} show running-difference images of STA/COR2 during 18:23$-$19:38 UT.
The CME enters the FOV of COR2 at 18:23:30 UT and propagates eastward with an angular width of $\sim$64$\degr$ (see Table~\ref{tab-2}).
The height evolution of the CME leading edge in the FOV of COR2 is plotted with green diamonds in Figure~\ref{fig3}(b).
A linear fitting results in an apparent speed of $\sim$1482 km s$^{-1}$.

\subsection{3D shapes of the prominence}
The eruptive prominence was not only observed by SDO/AIA as shown in Figure~\ref{fig4}, 
but also observed by the Sun Watcher using Active Pixel System detector and image processing (SWAP; \citealt{ber06}) in 174 {\AA} 
on board the PROBA 2 spacecraft with a larger FOV but a lower resolution than AIA, 
and the Solar Upper Transition Region Imager (SUTRI; \citealt{bai23}) onboard the Space Advanced Technology demonstration satellite (SATech-01).
SUTRI takes full-disk solar images at Ne {\sc vii} 465 {\AA} with a FOV of $\sim$41.6'$\times$41.6', a spatial resolution of $\sim$8$\arcsec$, and a normal cadence of 30 s.
The Ne {\sc vii} line is formed at $\sim$0.5 MK in the upper transition region (\citealt{tian17}).
Meanwhile, the Extreme-ultraviolet Imager (EUVI; \citealt{wu04}) on board STA detected the prominence in 195 and 304 {\AA} from another perspective (Figure~\ref{fig1}(b)).

   \begin{figure}
   \centering
   \includegraphics[width=0.8\textwidth, angle=0]{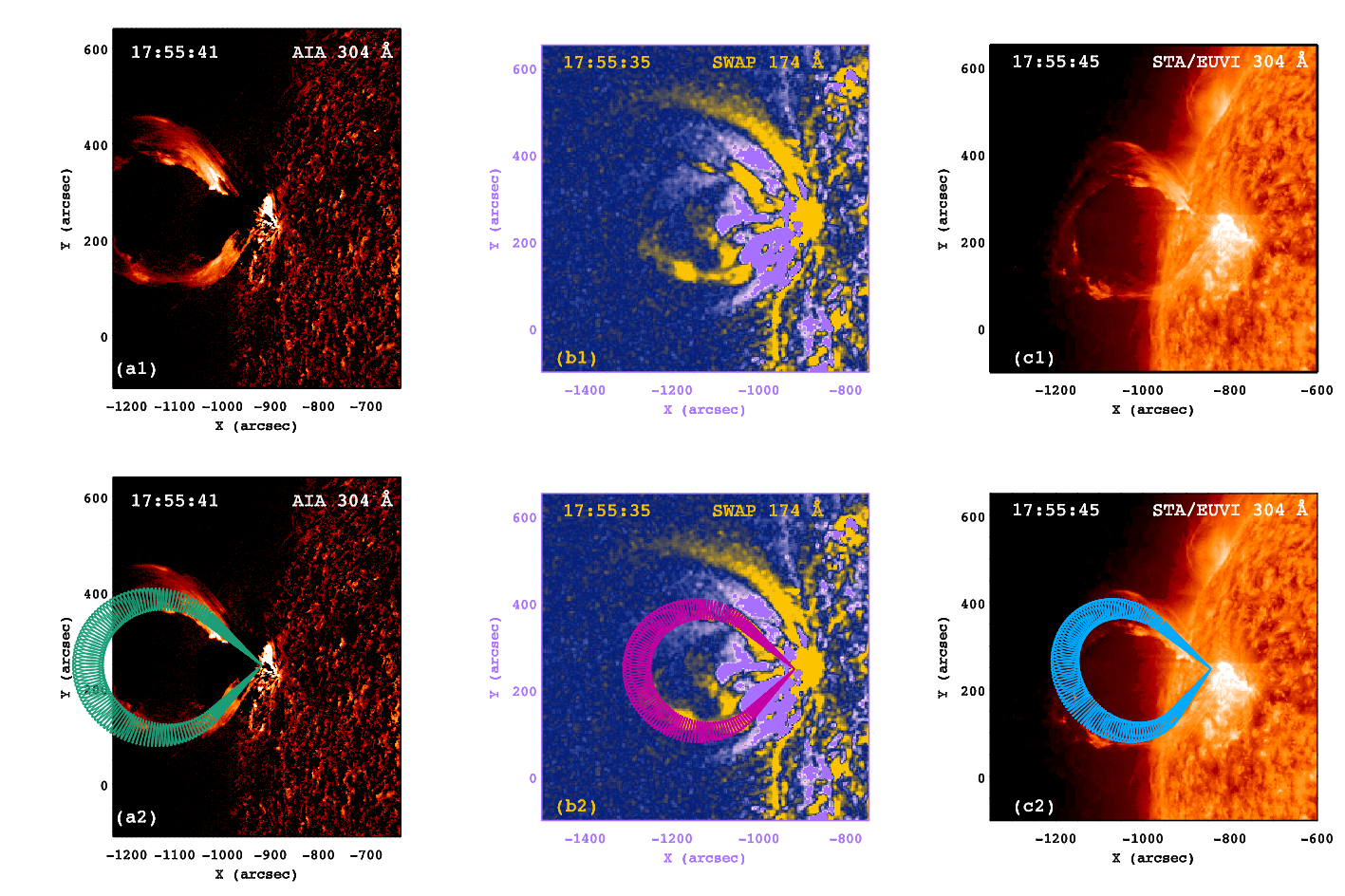}
   \caption{Top panels: the prominence observed by AIA 304 {\AA} (a1), SWAP 174 {\AA} (b1), and EUVI 304 {\AA} (c1) passbands around 17:55:40 UT.
   Bottom panels: the same images superposed with projections of the reconstructed flux rope (atrovirens, magenta, and blue dots).}
   \label{fig6}
   \end{figure}

In Figure~\ref{fig6}, the top panels show the prominence simultaneously observed by AIA 304 {\AA} (base-difference image), SWAP 174 {\AA} (base-difference image), 
and EUVI 304 {\AA} (original image) passbands around 17:55:40 UT. 
Due to the low cadence (10 minutes) of EUVI 304 {\AA} passband, this is the only time when the prominence is entirely visible in all instruments.
Owing to the smaller FOV of AIA than SWAP and EUVI, the whole prominence was captured by SWAP and EUVI, 
while the outermost part (i.e., apex) of the prominence was missed by AIA. It is obvious that the two legs are much brighter than the top of the prominence. 
In panel (c1), the prominence presents clear helical structure, implying that the magnetic fields supporting the prominence is most probably a flux rope. 
The bottom panels of Figure~\ref{fig6} show the same images, which are superposed with projections of the reconstructed flux rope (atrovirens, magenta, and blue dots) using the revised GCS model. 
The 3D reconstruction is performed by repeatedly adjusting the free parameters described in Section~\ref{sect:model}, while the source region location ($\phi=-84\degr$, $\theta=15\degr$) is fixed.
The best-fit model is subjectively judged when projections of the flux rope nicely match the prominence in EUV images. 
From Figure~\ref{fig6}(a2-c2), it is revealed that the fitting of the prominence using the revised GCS model is satisfactory. 
The derived parameters are: $h=150\arcsec$, $\alpha=45\degr$, $\kappa=0.087$ ($\delta=5\degr$), $\phi_1=0\degr$, $\theta_1=16\degr$, and $\gamma=20\degr$.
The height of leading ledge is $h_{\mathrm{LE}}=396\farcs6$, the edge-on width of the flux rope is $\omega_{\mathrm{EO}}=2\delta=10\degr$, 
and the face-on angular width is $\omega_{\mathrm{FO}}=2(\alpha+\delta)=100\degr$. 
The flux rope axis deviates from the local vertical direction by 16$\degr$ and the heliocentric distance ($h_{\mathrm HC}$) of the leading edge reaches $\sim$1.4 $R_\odot$.

   \begin{figure}
   \centering
   \includegraphics[width=0.8\textwidth, angle=0]{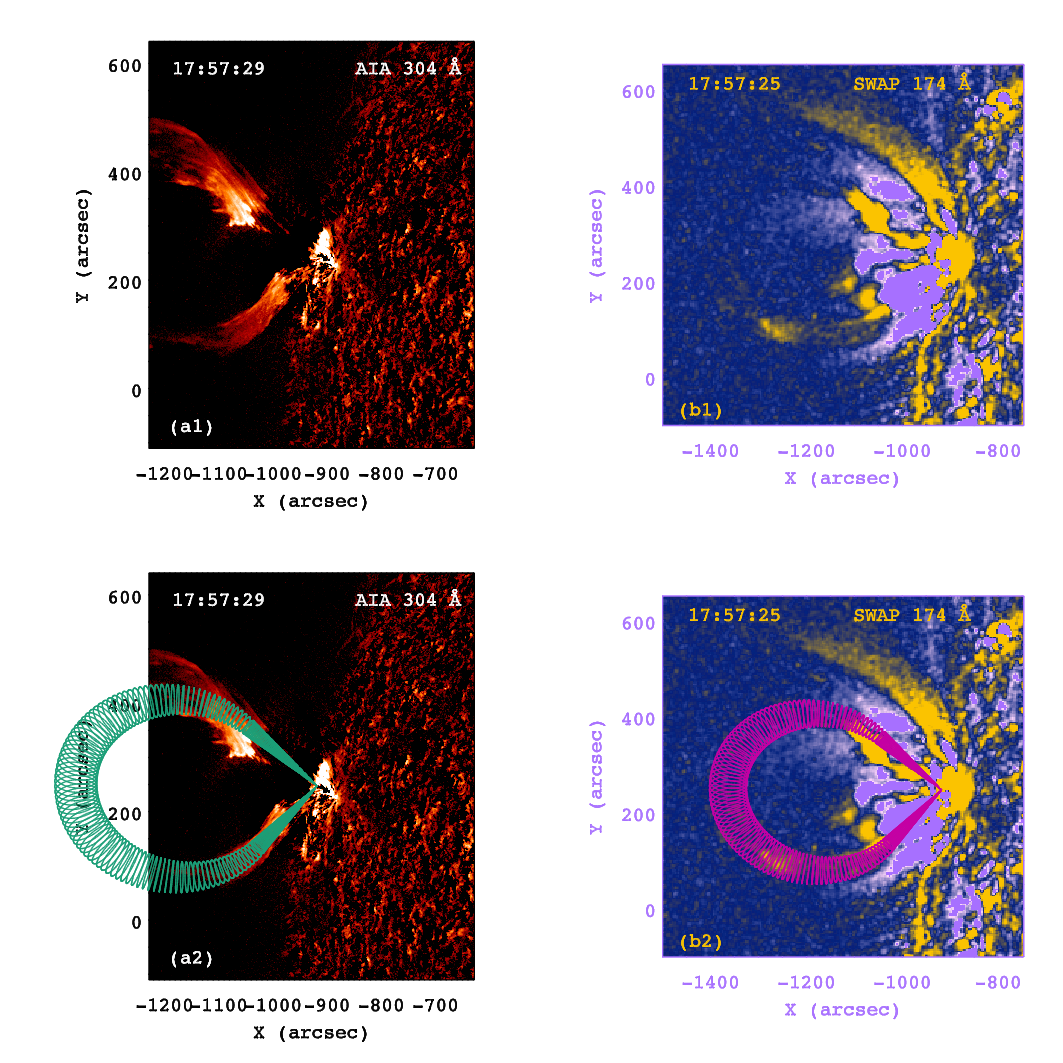}
   \caption{Top panels: the prominence observed by AIA 304 {\AA} and SWAP 174 {\AA} around 17:57:27 UT.
   Bottom panels: the same images overlaid with projections of reconstructed flux rope (atrovirens and magenta dots).}
   \label{fig7}
   \end{figure}

Although there is only one time of simultaneous observations of the prominence from multiple perspectives, 
3D reconstruction could still be conducted using observations of telescopes along the Sun-Earth connection (\citealt{the06}).
In Figure~\ref{fig7}, the top panels show the prominence observed by AIA 304 {\AA} and SWAP 174 {\AA} around 17:57:27 UT.
The prominence was fully visible in SWAP 174 {\AA} image at 17:57:25 UT, but was partly visible in AIA 304 {\AA} image at 17:57:29 UT.
The bottom panels show the same images overlaid with projections of reconstructed flux ropes (atrovirens and magenta dots).
Consistency between the shapes of prominence and flux ropes indicates that the fittings are still gratifying. The derived parameters are drawn in Figure~\ref{fig3}(c-d).

\begin{figure}
   \centering
   \includegraphics[width=0.8\textwidth, angle=0]{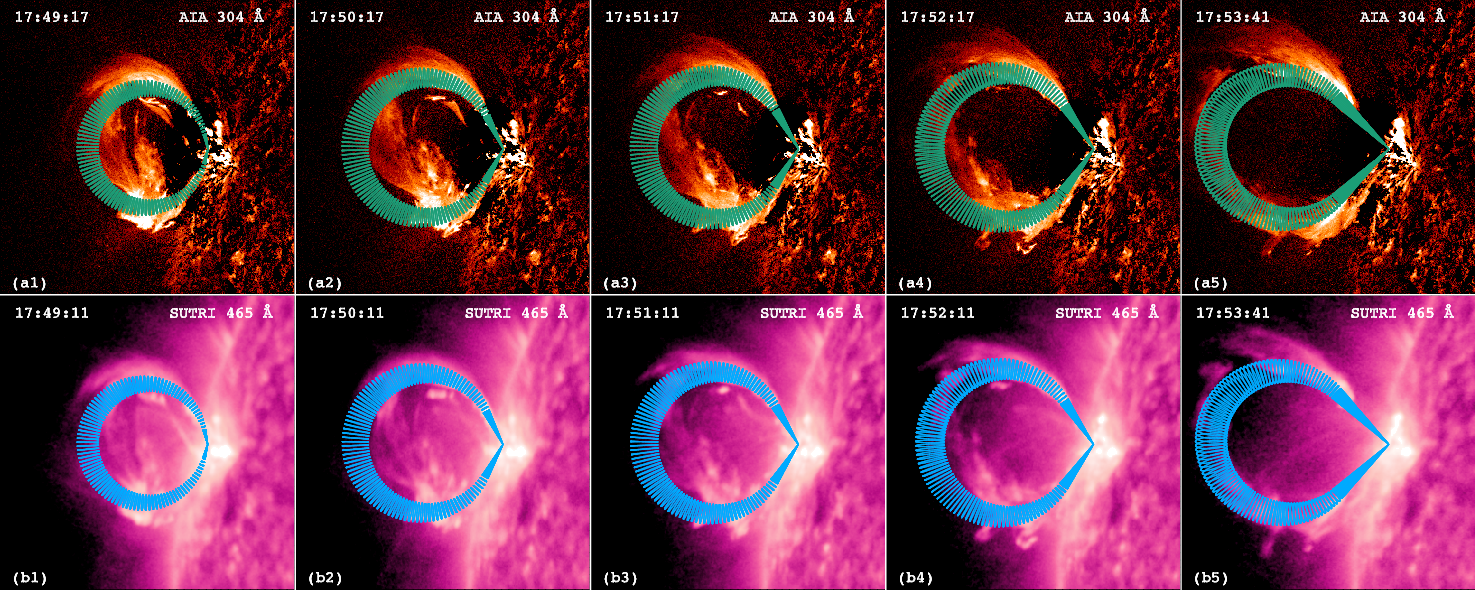}
   \caption{AIA 304 {\AA} images (a1-a5) and SUTRI 465 {\AA} images (b1-b5) superposed with projections of the reconstructed flux ropes (atrovirens and blue dots) during 17:49$-$17:53 UT.}
   \label{fig8}
   \end{figure}

Before 17:54:00 UT, the prominence rose gradually and was entirely recorded in AIA 304 {\AA} and SUTRI 465 {\AA} passbands.
Figure~\ref{fig8} shows 304 {\AA} images (a1-a5) and 465 {\AA} images (b1-b5) overlaid with projections of the reconstructed flux ropes (atrovirens and blue dots) during 17:49$-$17:53 UT.
The prominence looks like an ear and the two legs are much clearer than the top.
The reconstructed flux ropes coincide with the prominence much better at the legs than the top due to its irregular and asymmetric shape. The derived parameters are drawn in Figure~\ref{fig3}(c-d). 
Linear fittings of $h_{\mathrm{LE}}$ are separately performed during 17:49:17$-$17:52:17 UT and 17:53:30$-$17:57:30 UT, giving rise to true speeds of $\sim$246 and $\sim$708 km s$^{-1}$ of the erupting prominence.
Accordingly, the prominence was undergoing acceleration during its early phase of eruption (17:49$-$17:57 UT). 
In Figure~\ref{fig3}(b), time variation of $h_{\mathrm HC}$ is plotted with blue circles, which has the same trend as $h_{\mathrm{LE}}$.

The value of $\gamma$ increases from $0\degr$ to 30$\degr$, which is probably indicative of counterclockwise rotation of the prominence axis during eruption (\citealt{fan03, zhou20}).
The edge-on width $\omega_{\mathrm{EO}}$ keeps a constant of $\sim$10$\degr$. 
The face-on width $\omega_{\mathrm{FO}}$ decreases from $\sim$162$\degr$ to a minimum of $\sim$100$\degr$ around 17:53:45 UT and increases to $\sim$104$\degr$ around 17:57:25 UT.
The inclination angle $\theta_1$ increases slightly from 14$\degr$ to 16$\degr$, suggesting a southward deflection of the prominence.
The values of $\phi_1$ remain 0$\degr$, meaning that there is no longitudinal deflection.
In Table~\ref{tab-2}, the CPA of CME is 85$\degr$-88$\degr$, indicating a southward deflection of CME by 11$\degr$-14$\degr$.
In this regard, deflections of the prominence and related CME are accordant, which justifies the results of fitting using the revised GCS model.
Furthermore, the true speeds ($V_{3D}$) of CME are estimated to be 1653 and 1622 km s$^{-1}$ using the apparent speeds in the FOVs of LASCO/C2 and STA/COR2, which are very close to each other.
It is noted that the speed of CME (1637$\pm$15 km s$^{-1}$) is $\sim$2.3 times higher than that of prominence, implying continuing acceleration of the prominence between 17:57 UT and 18:23 UT.

\begin{table}
\begin{center}
\caption[]{Parameters of the CME Produced by the Prominence Eruption, Including the Apparent Speed ($V_{app}$), 
True Speed ($V_{3D}$), Central Position Angle (CPA), and Angular Width (AW).}\label{tab-2}
 \begin{tabular}{c|c|c|c|c}
  \hline\noalign{\smallskip}
Instrument &   $V_{app}$     & $V_{3D}$ & CPA & AW   \\
         &   (km $^{-1}$)    &  (km $^{-1}$)    &   (degree)   &   (degree)            \\
  \hline\noalign{\smallskip}
LASCO/C2 & 1644 & 1653 & 85 & 50 \\
STA/COR2 & 1482 & 1622 & 88 & 64 \\
  \noalign{\smallskip}\hline
\end{tabular}
\end{center}
\end{table}

\section{Summary and Discussion}
\label{sect:sum}

In this paper, the GCS model is slightly revised by introducing longitudinal and latitudinal deflections of prominences originating from ARs.
Subsequently, it is applied to the 3D reconstruction of an eruptive prominence in AR 13110, which produced an M1.7 class flare and a fast CME on 2022 September 23.
It is found that the prominence undergoes acceleration from $\sim$246 to $\sim$708 km s$^{-1}$. 
Meanwhile, the prominence experiences southward deflection by 14$\degr$-16$\degr$ without longitudinal deflection, suggesting that the prominence erupts non-radially.
Southward deflections of the prominence and associated CME are consistent, validating the results of fitting using the revised GCS model.
Besides, the true speed of the CME is calculated to be 1637$\pm$15 km s$^{-1}$, which is $\sim$2.3 times higher than that of prominence.
This is indicative of continuing acceleration of the prominence during which flare magnetic reconnection reaches maximum beneath the erupting prominence.
Hence, the reconstruction using the revised GCS model could successfully track a prominence in its early phase of evolution until $\sim$1.5 $R_\odot$, including acceleration and deflection.

Morphological reconstructions of prominences/filaments are abundant using stereoscopic observations in UV, EUV, and H$\alpha$ passbands from two or three viewpoints.
The triangulation method has been widely used to perform reconstructions of both quiescent and AR prominences (\citealt{li11, bi13, guo19}).
However, this method utilizes simultaneous images from two perspectives.
In the current study, there is only one moment ($\sim$17:55:45 UT) of observations from SDO/AIA and STA/EUVI when triangulation method is usable (Figure~\ref{fig6}).
On the contrary, the revised GCS model is at work even if there are observations from a single perspective (Figures~\ref{fig7},\ref{fig8}),
although more perspectives impose better constraints and have lower uncertainties.
This is particularly advantageous to the reconstruction of hot channels since routine observations in hot emission lines (such as 94, 131 {\AA}) with STEREO and SolO/EUI are still unavailable.
Calculations of the thermal energies of hot channels using this model will be the topic of our next paper.

Of course, there are limitations of the revised GCS model. Firstly, the model is applicable to AR prominences whose footpoints are close to each other, 
instead of quiescent prominences with much larger sizes and extensions. Secondly, the model is applicable to coherent, loop-like prominences, rather than those presenting irregular and ragged shapes.
Lastly, 3D reconstructions of prominences are severely constrained by the FOVs of solar telescopes working at UV, EUV, and H$\alpha$ wavelengths,
which is in contrast to the reconstructions of CMEs observed by coronagraphs with much larger FOVs.
In Figure~\ref{fig3}(b), the heliocentric distance of the flux rope leading edge reaches $\sim$1.5 $R_\odot$ at 17:57:25 UT, which is still blocked by the occulting disk of LASCO/C2.

With the advent of peak year of the twenty fifth solar cycle, large-scale solar eruptions are booming, which have sustained impact on near-Earth space environment.
Precise reconstructions of the shape and direction of eruptive prominences and the related CMEs will undoubtedly improve our ability of space weather forecast.
In the future, more case studies and statistical analysis are worthwhile using stereoscopic observations from spaceborne and ground-based telescopes, 
such as SDO/AIA, STEREO/EUVI, SolO/EUI, SWAP, SUTRI, the Chinese H$\alpha$ Solar Explorer (CHASE; \citealt{li22b}),
and the New Vacuum Solar Telescope (NVST; \citealt{liu14}).

\normalem
\begin{acknowledgements}
The authors appreciate Profs. Hui Tian and Hongqiang Song for helpful discussions.
SDO is a mission of NASA\rq{}s Living With a Star Program. AIA data are courtesy of the NASA/SDO science teams.
SUTRI is a collaborative project conducted by the National Astronomical Observatories of CAS, Peking University, Tongji University, 
Xi'an Institute of Optics and Precision Mechanics of CAS and the Innovation Academy for Microsatellites of CAS.
This work is supported by the National Key R\&D Program of China 2022YFF0503003 (2022YFF0503000), 2021YFA1600500 (2021YFA1600502), 
and Yunnan Key Laboratory of Solar Physics and Space Science under the number YNSPCC202206.
\end{acknowledgements}

\label{lastpage}
\clearpage

\end{document}